# Schrödinger Equation weakly attractive $1/r^2$ Potential Eigenfunctions

Philip E. Bloomfield

(Drexel University School of Biomedical Engineering, Science and Allied Health Professions)

*Abstract*: Bound state solutions of the Schrödinger Equation for the $-\lambda/r^2$ potential have been presented recently for both the weak and strong coupling cases. However, Shortley in 1931 and Landau and Lifshitz in 1958 claimed that no bound state solutions exist for the weak coupling case when $0 < 2m\lambda/\hbar^2 \leq (\ell + \frac{1}{2})^2$. We demonstrate that one bound state solution can exist for each angular momentum state $\ell$, and that a complete orthogonal set of continuum eigenfunctions orthogonal to the bound state eigenfunction can be constructed when $\ell(\ell+1) < 2m\lambda/\hbar^2 \leq (\ell + \frac{1}{2})^2$. We show that Shortley's argument is spurious due to his neglecting a boundary term arising from the momentum operator and that the Landau and Lifshitz claim is based on a restrictive fitting of the exterior solution to an interior spherical well. Instead, to each weak coupling, $\lambda$, we find a unique interior well strength $V_{in} = -\lambda' r_0^{-2}$ which yields a finite bound state. In particular for the $\ell = 0$ case the well strength is given by: $2m\lambda'/\hbar^2 \approx 1.3734 + 2.2265*(\frac{1}{4} - 2m\lambda/\hbar^2)^{\frac{1}{2}}$.

## I. Introduction

Recently, the $-\lambda/r^2$ potential Schrödinger equation has been studied as a model relevant to the three-body problem in nuclear physics[1,2,3] and also to point dipole interactions in molecular physics.[3,4] Included in these studies were both the weak coupling case and the strong coupling case, $2m\lambda/\hbar^2 > (\ell + \frac{1}{2})^2$, where an infinite spectrum of discrete bound states exists with energies extending from 0 to $-\infty$. The eigenstate behaviors of the weak and strong $-\lambda/r^2$ potential were used to model second order phase transitions.[3,5] In Sections II and III of the present study the weakly attractive $1/r^2$ potential Schrödinger equation is reviewed and errors in Shortley's[6] investigation are pointed out and corrected. In Section IV it is shown that when $(\ell+ ½)^2 \geq 2m\lambda/\hbar^2 > \ell(\ell+ 1)$ a single bound state exists. Section V reviews and extends the Landau and Lifshitz[7] calculation for the weak coupling case to reveal the existence of a single allowed bound state. In Section VI the matching of the interior spherical well potential, $-\lambda'/r_0^2$, solution to the exterior solution implies a unique interior parameter $\lambda'$ determined by the exterior parameter, $\lambda$. Section VII presents the set of continuum wave functions which are orthogonal to the single allowed bound state. In Section VIII the spectrum corresponding to the strong coupling potential is presented. Finally, the summary and conclusions are contained in Section IX.

## II.     Review of Shortley's Work

In 1931 Shortley[6] studied the non-relativistic Schrödinger equation

$$-\frac{\hbar}{2m}\nabla^2\psi + V\psi = i\hbar\partial_t\psi, \qquad V = -\lambda/r^2, \qquad \lambda > 0 \qquad (1)$$

and found a number of surprising results.

By making a series of substitutions Eq. (1) can be brought into the form considered by Shortley.

The substitution

$$\psi = r^{-1}\chi(r)Y_l^m(cos\theta,\phi)e^{-iEt/\hbar} \qquad (2)$$

brings Eq. (1) into the Hamiltonian form

$$H\chi = -\frac{\hbar^2}{2m}[\partial_r^2 - \ell(\ell+1)r^{-2}]\chi + V\chi = E\chi. \qquad (3)$$

Then the substitution

$$\Gamma = \gamma - \ell(\ell + 1) \qquad \text{where} \qquad \gamma = 2m\lambda/\hbar^2 \qquad (4)$$

yields for the $V$ of Eq. (1)

$$\partial_r^2 \chi + (\tfrac{2m}{\hbar^2} E + \Gamma r^{-2})\chi = 0 \qquad (5)$$

while for $E < 0$ the change of variable

$$\rho = (-2mE/\hbar^2)^{1/2} r \qquad (6)$$

brings Eq. (5) into the form considered by Shortley

$$\partial_\rho^2 \chi + (-1 + \Gamma \rho^{-2})\chi = 0. \qquad (7)$$

For $\Gamma > \tfrac{1}{4}$ (hypercritical potential) Shortley found quadratically integrable bound state solutions of the form

$$\chi = \rho^{1/2} K_{i|\nu|}(\rho) \qquad (8)$$

where $|\nu| = (\Gamma - \tfrac{1}{4})^{1/2}$. See Section VIII for a discussion of the hypercritical potential solutions.

However, for $\Gamma \leq \tfrac{1}{4}$ where

$$\nu = (\tfrac{1}{4} - \Gamma)^{1/2} = [(\ell + \tfrac{1}{2})^2 - \gamma]^{1/2}, \qquad (9)$$

Shortley found for each negative eigenvalue that the corresponding eigenfunction is positive definite; hence, two different eigenfunctions cannot be orthogonal. On the basis of non-orthogonality and completeness, Shortley allows the existence of at most one negative energy solution for $\Gamma \leq \tfrac{1}{4}$:

$$\chi = \rho^{1/2} K_\nu(\rho) \tag{10}$$

where $K_\nu(\rho)$ is a Bessel function.[8,9]

However, Shortley rejected this single level because of an (incorrect) adjoint operator argument and thereby concluded that all negative eigenvalues of the Hamiltonian can be eliminated for $\Gamma \leq \frac{1}{4}$. Our examination shows that his argument is flawed.

Shortley expressed the Hamiltonian operator as the product of two operators, $\bar{A}$ and $A$, that are related to the radial momentum operator so that $H = \bar{A}A$:

$$A = (2m)^{-1/2} p_r - i\alpha r^{-1}, \qquad \bar{A} = (2m)^{-1/2} p_r + i\alpha r^{-1}. \tag{11}$$

The parameter $\alpha$ is determined in Eq. (15) and the radial momentum operator is expressed as

$$p_r \psi = -i\hbar r^{-1} \partial_r (r\psi). \tag{12}$$

The radial momentum operator obeys the relations:

$$[p_r, r^{-1}]\psi = i\hbar r^{-2}\psi, \qquad p_r^2 \psi = -\hbar^2 (r^{-1}\partial_r r)(r^{-1}\partial_r r)\psi = -\hbar^2 r^{-1} \partial_r^2 (r\psi). \tag{13}$$

Thus the Hamiltonian $H$ can be expressed as:

$$H\psi = (2m)^{-1}(p_r^2 - \hbar^2 \Gamma r^{-2})\psi = \bar{A}A\psi \tag{14}$$

if $\alpha$ is real and satisfies Eq. (15) which is obtained by calculating $\bar{A}A$ from Eq. (11):

$$\alpha^2 + \hbar\alpha(2m)^{-1/2} = -\hbar^2(2m)^{-1}\Gamma. \tag{15}$$

The quadratic Eq. (15) has solution, $\alpha = (\pm \nu - \frac{1}{2}) \hbar (2m)^{-\frac{1}{2}}$ showing real $\alpha$ only for $\Gamma \leq \frac{1}{4}$.

Shortley[6] presented the following Eq. (16) from which he (incorrectly) deduced that $\bar{A}$ is the adjoint of A and consequently the energy $E$ cannot be negative.

$$" E<\psi|\psi> = <\psi|H\,\psi> = <\psi|\bar{A}A\,\psi> = <A\psi|A\psi> = \int_0^\infty r^2 |A\psi|^2 dr > 0 \quad ." \quad (16)$$

From this he concluded that $E > 0$ and therefore no bound states exist for $\Gamma \leq \frac{1}{4}$. This argument was quoted verbatim and called[10] an elegant proof of Hardy's inequality. Shortley's argument will next shown to be at fault.

### III. Critical Examination of Shortley's Calculation

Let us to examine Shortley's calculation keeping in mind von Neuman's remark[11] to carefully consider contributions from boundary terms when dealing with the momentum operator. Upon integrating by parts, we find:

$$E<\psi|\psi> = <\psi|H\,\psi> = \int_0^\infty r^2 \psi^* \bar{A} A \psi \, dr = <\psi|\bar{A}A\,\psi> =$$

$$= <A\psi|A\psi> - \frac{\hbar^2}{2m}[\chi^*(\partial_r + \frac{\nu - \frac{1}{2}}{r})\chi]\Big|_0^\infty \quad (17)$$

Examination of the boundary term shows that it is non-zero and hence Shortley's argument fails.

We chose the root of Eq. (15), $\alpha = (\nu - \frac{1}{2}) \hbar (2m)^{-\frac{1}{2}}$, so that the integral $<A\psi|A\psi>$ and the boundary term are finite. The case $\nu = 0$ and the root, $\alpha = -(\nu + \frac{1}{2})\hbar(2m)^{-\frac{1}{2}}$, give infinite values for both the boundary term and the integral $<A\psi|A\psi>$.

Substituting Eq. (10) into the boundary term of Eq. (17) yields:[12]

$$[\chi^*(\partial_r + \frac{\nu - \frac{1}{2}}{r})\chi]\Big|_0^\infty = (-2mE/\hbar^2)^{1/2}[\chi^*(\partial_\rho + \frac{\nu - \frac{1}{2}}{\rho})\chi]\Big|_0^\infty = \frac{1}{2}\pi(-2mE/\hbar^2)^{1/2}\csc\pi\nu \quad (18)$$

Note that the calculation in Eq. (18) utilizes the first two terms in the expression for χ in[12] while Eq. (7) is satisfied by all four terms in[12].

The existence of the boundary term in Eq. (17) shows that Ā is not the adjoint of A for the operators of Eq. (11). The left side of Eq. (17) must be negative for a bound state to exist. Eq. (18) implies that a necessary condition for this is

$$\sin\pi\nu > 0 \quad (19)$$

which implies that $0 < \nu < 1$. In Section IV the normalization integral $<\Psi/\Psi>$ is evaluated and shown to be positive for $0 \leq \nu < 1$. Thus bound states can exist in the subcritical region ($\Gamma \leq \frac{1}{4}$).

### IV.    Conditions Allowing Bound State Existence -- Orthonormality Calculation

Next the values of ν (or γ) which allow a bound state are found. This is accomplished by computing the normalization integral, $<\Psi/\Psi>$ in terms of the eigenfunction and its derivative with respect to energy and radius. Integrating the result of multiplying Eq. (3) for $E = E_2$ by the complex conjugate wave function for $E = E_1$ and subtracting the corresponding conjugated equation for $E_2$ and $E_1$ transposed, there results:

$$0 = 2m\hbar^{-2}(E_1 - E_2)\int_0^\infty \chi_1^*\chi_2 dr = (\chi_1^*\partial_r\chi_2 - \chi_2^*\partial_r\chi_1)\Big|_0^\infty. \quad (20)$$

If $E_2 \neq E_1$, then the left hand side of Eq. (20) states the orthogonality condition. However the evaluation of the right hand side of Eq. (20), which expresses the Hamiltonian operator's Hermiticity, shows that for $\Gamma < ¼$ that[12] $(-E_1/-E_2)^\nu = 1$; i.e., $\ln(|E_2|/|E_1|) = 0$. This implies that there can be only one negative eigenvalue.

The normalization integral can be calculated[13] by dividing through Eq. (20) by $(E_2-E_1)$ and using L'Hospital's rule:

$$2m\hbar^{-2}\int_0^\infty \chi^2 dr = -(\chi \partial^2_{E,r}\chi - \partial_E \chi \, \partial_r \chi)|_0^\infty . \tag{21}$$

Evaluating Eq. (21) by means of Eq. (10) gives[12] for $\nu \neq 0$:

$$2mE\hbar^{-2}<\psi/\psi> \, = \, 2mE\hbar^{-2}\int_0^\infty \chi^2 dr = -\tfrac{1}{2}\pi\nu(-2mE\hbar^{-2})^{1/2}\csc \pi\nu \, < 0. \tag{22}$$

The Eq. (22) $\nu = 0$ limit yields the same result as the $\nu = 0$ ($\Gamma = ¼$) evaluation of Eq. (21):

$$-½\,(-2mE\hbar^{-2})^{½}.$$

Finally, from Eqs. (17), (18), and (22) the inequality of Eq. (23) results:

$$0 < 2m\hbar^{-2}\int_0^\infty r^2 |A\psi|^2 dr = \tfrac{1}{2}\pi(-2mE\hbar^{-2})^{1/2}(1-\nu)\csc \pi\nu , \tag{23}$$

which in agreement with Eq. (19) requires $\nu < 1$. Also Eq. (22) shows that the existence of a bound state requires $0 \leq \nu < 1$ which from Eq. (9) implies that $\Gamma > -¾$. However, there is a restriction on the value of $\nu$ due to the fact that the solution $\psi = \chi/r$, with $\chi$ as given in Eq. (10) does not satisfy the Schrödinger equation if $\nu \geq ½$ (i.e., if $\Gamma < 0$).

To show this, integrate the Schrödinger equation, $(\nabla^2 + \Gamma/r^2 + 2mE/\hbar^2)\psi = 0$, within a spherical volume containing the origin and then let the volume shrink toward zero (where[12] $\psi \sim r^{-\frac{1}{2}-\nu}$ ):

$$\int d^3r \nabla^2 \psi = \oiint d\vec{S}\cdot\vec{\nabla}\psi = \lim_{r\to 0}\left[-4\pi(\tfrac{1}{2}+\nu)r^{1/2-\nu}\right]. \tag{24}$$

If this expression differs from zero, then $\psi$ does not satisfy the Schrödinger equation for the $r^{-2}$ potential.[14] Thus the bound state solution only exists if $\nu > 0$ from Eqs. (19) and (23) and $\nu < \tfrac{1}{2}$ from setting Eq. (24) to zero.

Thus a single bound state can exist if:

$$0 \leq \nu < \tfrac{1}{2} \quad \text{or} \quad (\ell+\tfrac{1}{2})^2 \geq \gamma > \ell(\ell+1) \tag{25}$$

If $\nu$ is imaginary ($\Gamma > \tfrac{1}{4}$; i.e., $(\ell+\tfrac{1}{2})^2 < \gamma$), then an infinite sequence of bound states arises, as discussed in Section VIII. As already shown if $\nu \geq \tfrac{1}{2}$, then $\ell(\ell+1) \geq \gamma$ and the eigenfunction of Eq. (10) does not satisfy the Schrödinger equation and furthermore the $\ell = 0$ state would imply a repulsive potential. Some authors[15,16] have not taken into account the Eq. (24) condition and have concluded that repulsive potentials in the Schrödinger equation can yield bound states.

## V. Critical Review of Landau and Lifshitz (L.L.) Treatment

The Landau and Lifshitz[7] proof of the non-existence of $\ell = 0$ bound states for the $1/r^2$ potential depends on their choice of interior potential. Traditionally when the Schrödinger equation is expanded in a Frobenius series about a singular point and two roots of the indicial equation are found with both of them yielding divergent behavior of $\psi$ at the origin, the solution which becomes infinite less rapidly is kept.[7,17] For small $r$ the Bessel function in Eq. (10) can be approximated by the first two terms of a power series. Landau and Lifshitz (L.L.)[7] wrote:

$$\psi \sim Ar^{s_+} + Br^{s_-}, \qquad s_\pm = -\tfrac{1}{2} \pm \nu, \qquad \nu \neq 0 \qquad (26)$$

For $r < r_0$, L.L. replaced the $r^{-2}$ potential by a particular interior potential $V_0$ = constant = $V(r_0)$ and boundary conditions were matched. Finally L.L. let $r_0 \to 0$ and concluded that if $\Gamma < \tfrac{1}{4}$ and $V = -\lambda r^{-2}$ in all space, then[7] "there are no negative energy levels." On the spherical surface where $r = r_0$, L.L. matched ($r_0$ times) the logarithmic derivative of $r\psi$ outside a small sphere of radius $r_0$,

$$r_0 \partial_r \ln(r\psi) = [A(s_+ + 1)r_0^{s_+} + B(s_- + 1)r_0^{s_-}][Ar_0^{s_+} + Br_0^{s_-}]^{-1}, \qquad (27)$$

with ($r_0$ times) the logarithmic derivative of $r\psi$ inside the sphere where for $\ell = 0$, $V = -\lambda' r_0^{-2}$ and $r\psi_{in} = \sin(\Lambda^{1/2} r/r_0)$:

$$r\partial_r \ln(r\psi_{in})|^{r=r_0} = \Lambda^{1/2} \cot \Lambda^{1/2} \qquad (28)$$

$$\Lambda = \gamma' + 2mE\,r_0^2 \hbar^{-2} \qquad \Lambda(r_0 \to 0) = \gamma' = 2m\lambda'/\hbar^2 \qquad (29)$$

From Eqs. (26), (27) and (28) there follows:

$$\frac{(\tfrac{1}{2}+\nu)A\,r_0^\nu + (\tfrac{1}{2}-\nu)B r_0^{-\nu}}{A r_0^\nu + B r_0^{-\nu}} = \Lambda^{1/2} \cot \Lambda^{1/2} \qquad (30)$$

$$[\tfrac{1}{2} - \nu - \Lambda^{1/2}\cot\Lambda^{1/2}]B/A = -[\tfrac{1}{2} + \nu - \Lambda^{1/2}\cot\Lambda^{1/2}]r_0^{2\nu}. \qquad (31)$$

L.L. conclude that since the right hand side of Eq. (31) approaches zero as $r_0 \to 0$ then $B/A \to 0$ and $\psi \to Ar^{s_+}$. Such a wave function has no oscillations and hence L.L. conclude that $\psi = Ar^{s_+}$ corresponds to the ground state. Since Eq. (26) is the exact (non-normalizable) solution of Eq. (5) for $E = 0$, they concluded that for their choice of interior potential, no $E < 0$ solutions exist.

In order that the exterior solution vanish at $r = \infty$ the particular Bessel function given in Eq. (10) must be chosen. Then the coefficients A and B in Eq. (26) must have the particular relation to each other that is given in Eq. (32). Comparison of the coefficients of $r^{½+v}$ and $r^{½-v}$ in Eq. (26) and in [12] yields:

$$B/A = -\Gamma(1+v)\{[-mE/(2\hbar^2)]^v \Gamma(1-v)\}^{-1} \qquad (32)$$

which does not vanish except in the special case, $E = -\infty$.[18]

Consistency of Eq. (31) in the limit $r_0 \rightarrow 0$ can be achieved for finite $E < 0$ if one considers an alternative interior potential. From Eq. (31) instead of choosing $B = 0$, the contents of the square bracket multiplying B/A can tend to zero; i.e. an interior solution $\psi_{in}$ having $(r\partial_r \ln r\psi)_{in}$ equaling $(½ - v)$ in the limit $r_0 \rightarrow 0$ suffices.[19] Taking the limit $r_0 \rightarrow 0$ in Eq. (30) yields $(½ - v) = \gamma'^{½} \cot \gamma'^{½}$. The function $K_v(\rho)$ in Eq. (10) has no oscillations for all $E < 0$. Thus, there is only one bound (ground) state for a $\gamma'$ determined by $\gamma$ as shown in Section VI.

## VI. Calculation of the ℓ = 0 interior potential, $-\lambda'/r_0^2$, in the limit $r_0 \rightarrow 0$

The arguments presented in Sections III-V show that the Eq. (10) Schrödinger equation eigenfunction describes one finite energy bound state for the subcritical $\gamma/r^2$ potential. The exterior solution (valid as $r \rightarrow \infty$) can be matched via Eq. (30) to the interior Schrödinger equation eigenfunction for a spherical square potential having a strength parameter $\gamma'$ dependent on the exterior potential parameter $\gamma$.

Section V revealed the limit $r_0 \to 0$ conditional equality (independent of *E*):

$$0 < (½ - \nu) = [½ - (¼ - \gamma)^{½}] = r\partial_r \ln(r\psi_{in})|_{r_0 \to 0} = \gamma'^{½} \cot\gamma'^{½} \leq ½ \tag{33}$$

Equation (33) shows that the quantities, (½ - $\nu$) and $\gamma'^{½}\cot\gamma'^{½}$, can be plotted on a common x-axis after being calculated from the variables $\gamma$ and $\gamma'$, respectively. Plotting the corresponding $\gamma$ and $\gamma'$ parameters on the y-axis shows that the exterior parameter $\gamma$ implies a single interior parameter $\gamma'$. that can be approximately fit to a linear function or more precisely to a quadratic function of (½ − $\nu$) as .shown in Eqs. (34a) and (34b), respectively, and displayed in Fig. 1.

$$\gamma' \approx 2.4867 - 2.2265*(½ - \nu), \tag{34a}$$

$$\gamma' \approx -0.4520*(½ - \nu)^2 - 1.9905*(½ - \nu) + 2.4671. \tag{34b}$$

**Ranges of the Variables *γ* and *γ′***

The solutions to the two limiting equalities in Eq. (33) yield:-

$$0 < \gamma'^{½}\cot\gamma'^{½} \leq ½ \quad \Rightarrow \quad 1.570796 \approx ½\pi > \gamma'^{½} \geq 1.16556 \tag{35}$$

$$2.46740 > \gamma' \geq 1.35853, \qquad 0 < \gamma \leq ¼. \tag{36}$$

The two limiting values of $\gamma'^{½}$ were determined from the iteration of $\gamma'^{½}\cot(\gamma'^{½}) = 0$ and ½.

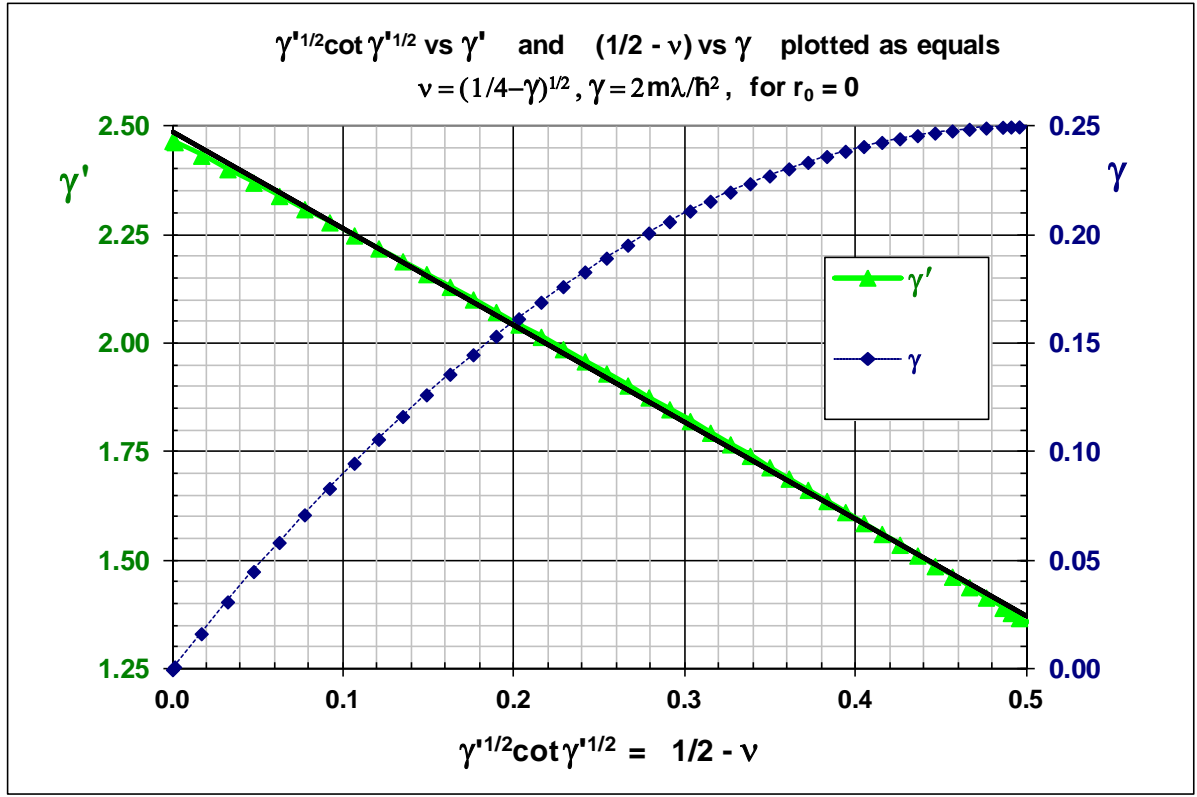

**Fig. 1**. The interior spherical well parameter $\gamma'$ determined by the exterior potential parameter $\gamma$. Approximate linear fit of $\gamma'$ vs $(½ − \nu)$ is shown and described in Eq. (34a).

## Two Special Cases

**No exterior potential:** $\nu = ½$ ($\gamma = 0$) and $\chi_{ext}$ represents a particle trapped in the well.

$$\chi_{ext} = \exp(-\rho) \tag{37}$$

Thus as $r_0 \to 0$, $(r_0 \partial_r \chi)/\chi = [-(-2mE/\hbar^2)^{½} r_0] \to 0 = \gamma'^{½} \cot \gamma'^{½} \Rightarrow \gamma'^{½} = ½\pi$ and so $\gamma' \approx 2.46740$.

**The transitional case:** $\nu = 0$ ($\gamma = ¼$) and[12] for small $r$,

$$\chi_{ext} \approx \rho^{½}[\psi(1) - \ln(½\rho)]. \tag{38}$$

Taking the limit $\nu \to 0$ in Eq. (30) yields: $½ = \Lambda^{½} \cot \Lambda^{½}$.

Thus, $(r_0 \partial_r \chi)/\chi = \{½[\psi(1) - \ln(½\rho)] - 1\}/\{\psi(1) - \ln(½\rho)\} = \gamma'^{½} \cot \gamma'^{½}$ and as $r_0 \to 0$,

[$\gamma'^{1/2}\cot\gamma'^{1/2}$] → ½ which was numerically solved by iteration to yield $\gamma' \approx 1.35853$.

The two limiting solutions[12] at $\nu = ½$ ($\gamma = 0$, $\gamma' = 2.46740$) and at $\nu = 0$ ($\gamma = ¼$, $\gamma' = 1.35853$) are seen here in the curves of Fig. 1. Note that in Ref.[5] the $\gamma = ¼$, $\gamma' \approx 1.36$ case was presented.

## VII. Continuum Wave Functions Complete the Hermitian Operator's Domain ---

Not only must the bound state eigenfunctions be mutually orthogonal; but continuum states which are orthogonal to the bound state(s) must be chosen. Otherwise the determination of the domain of a Hermitian (symmetric) operator has not been completed. Of the two independent solutions to Eq. (3) for $E > 0$, Shortley[7] kept $\chi = \rho^{1/2}J_\nu(\rho)$ and rejected $\chi = \rho^{1/2}Y_\nu(\rho)$ because he believed that "the normalization integral and the Hamiltonian integral do not converge at the origin for the Neumann function solutions, $\rho^{1/2}Y_\nu(\rho)$" and thus a bound state could be decomposed in terms of the (non-orthogonalized $\rho^{1/2}J_\nu(\rho)$) continuum states.

However for $\nu^2 < ½$ (which is the case since $\nu < ½$, subcritical and $\nu = i|\nu|$ hypercricial) the normalization and total Hamiltonian integrals converge near the origin for both solutions, even though the potential and kinetic energy contributions to the energy integral separately diverge. Thus we can write the continuum wave function as a sum of the $J_\nu$ and $Y_\nu$ Bessel functions:

$$\chi_1 = (k_1 r)^{1/2}[A_1 J_\nu(k_1 r) + B_1 Y_\nu(k_1 r)], \quad \text{where } k_1 = (2mE_1/\hbar^2)^{1/2} \quad \text{and } E = E_1 > 0. \quad (39)$$

Then taking $E = -|E_0|$ for the one bound state and $k_0 = (2m|E_0|/\hbar^2)^{1/2}$, the orthogonality (Hermiticity) condition can be calculated[20] from Eq. (20). This yields a relationship between $A_1$ and $B_1$; viz.

$$A_1/B_1 = \begin{cases} [(k_0/k_1)^{2\nu} - \cos \pi\nu]\csc \pi\nu, & \nu \neq 0 \\ \frac{2}{\pi}\ln(k_0/k_1), & \nu = 0 \end{cases} \tag{40}$$

Note that for the $E = 0$ continuum state, $k_1 = 0$ which implies $B_1 = 0$ as remarked in Section V.

## VIII. Hypercritical Singular Potential

The problem of highly singular potentials in quantum mechanics was examined by Case[21] who considered strongly attractive potentials ($r^{-n}$, n ≥ 2) in the Schrödinger equation as well as the Coulomb potential in the relativistic Dirac and spin zero and one equations. Case considered the bound state spectrum associated with the strongly attractive $-\gamma/r^2$ potential (for ¼ < γ and $\ell = 0$) and noted that both solutions of Schrödinger's equation have essentially the same behavior for $r \to 0$. The difficulty in deciding between solutions can be resolved if use is made of von Neumann's criterion that the set of eigenfunctions be orthonormal. Using this criterion, Case found a discrete bound state spectrum which depends on an arbitrary parameter B, the phase of the wave function (as $r \to 0$). He pointed out that B is not bounded from below and that if all the eigenfunctions have the same phase (as $r \to 0$) orthogonality will be obtained. Case notes the principal difference between singular and non-singular potentials: in the latter case the solutions to Schrödinger's equation, subject to the condition of quadratic integrability, form a complete orthonormal set; in the singular case the solutions are too numerous and overcomplete as had already been noted by Shortley[7]. Some other parameter in addition to that occurring in the potential is needed to completely specify the quantized level scheme. Case found it convenient to take the phase B at the origin as the parameter. Morse and Feshbach[22] also considered the Schrödinger equation for the $r^{-2}$ potential and $\ell = 0$. They did not find any quantization for small coupling constants; while for large coupling constants (¼ < γ in Eq. (5)) they found that the requirement that the bound states be orthogonal imposes a "strange quantization, not uniquely fixing the levels…but just fixing levels relative each other." They give the energy eigenvalue formula,

$$E_n = -|E_0|\exp[2\pi n(\gamma - \tfrac{1}{4})^{-1/2}], \ n = 0, \pm 1, \pm 2, \ldots, \tag{41}$$

where $E_0$ is an arbitrary parameter; this is in essential agreement with Case: a bound state spectrum extending from $-\infty$ to an accumulation point at zero energy. Note that Morse and Feshbach gave the wave function near the origin as:

$$\chi \sim \rho^{1/2} \operatorname{csch}(\pi|\nu|) \sin(|\nu| \ln(\tfrac{1}{2}\rho) - \Phi_{|\nu|}) \text{ where } \Phi_{|\nu|} = \arg\Gamma(1+i|\nu|). \tag{42}$$

We note that as $|\nu| \to 0$ ($\gamma \to \tfrac{1}{4}$) the wave function of Eq. (42) reduces to the limiting weak field case $\nu \to 0$ ($\gamma \to \tfrac{1}{4}$) of Eq. (38) and the energy eigenvalues of Eq. (41) reduce to the three values, $-|E_0|, 0, -\infty$.

## IX. Summary and Conclusions

Arguments were presented disproving the claims of Shortley and Landau and Lifshitz that no bound state can exist for a weak $-\lambda/r^2$ potential, $0 < 2m\lambda/\hbar^2 \leq \tfrac{1}{4}$, in the Schrödinger equation. It was shown that one negative energy level is allowed for each $\ell$ if $(\ell + \tfrac{1}{2})^2 \geq 2m\lambda/\hbar^2 > \ell(\ell + 1)$, the latter inequality is necessitated for compliance with the Schrödinger equation.[14,15] That is, delta-function behavior of the wave function at the origin is rejected.

Furthermore we derived criteria for matching the exterior solution to an interior spherical well. For each allowed $\lambda$ a unique interior well strength parameter $\lambda'$ yields a finite bound state. The set of continuum states was made orthogonal to the bound state, a necessary condition for the Hamiltonian to be a Hermitian operator. For the $\ell = 0$ case it was shown that as $2m\lambda/\hbar^2$ becomes infinitesimally larger than ¼, one state has the value, $-|E_0|$, corresponding to the $2m\lambda/\hbar^2 = ¼$ situation. Also for $2m\lambda/\hbar^2 > ¼$, an infinite sequence of bound levels arises near the continuum at $E = 0$ and another infinite sequence of bound levels appears near $E = -\infty$. Landau and Lifshitz[7,18] call this latter process (the existence of a bound state at $E = -\infty$) "fall to the center".

[9] This Bessel function vanishes exponentially as $\rho \to \infty$: $\rho^{1/2} K_\nu(\rho) \to (\frac{\pi}{2})^{1/2} \exp(-\rho) = \rho^{1/2} K_{1/2}(\rho)$; see Ref [8], Sect. 9.7. From Ref [8], Sect. 9.6, $K_\nu(\rho) = \frac{1}{2} \pi \csc \pi \nu [I_{-\nu}(\rho) - I_\nu(\rho)]$,

$I_\nu(\rho) = \sum_0^\infty (\frac{1}{2}\rho)^{\nu+2n} [n! \Gamma(\nu+n+1)]^{-1}$, $K_0(\rho) = -I_0(\rho) \ln(\frac{1}{2}\rho) + \sum_0^\infty (\frac{1}{2}\rho)^{2n} (n!)^{-2} \psi(n+1)$,

where from Ref [8], Sect. 6.3 $\psi(n)$ is the digamma function.

[10] H. Kalf, U.W. Schmincke, J. Walter, and R. Wüst, "On the Spectral Theory of Schrödinger and Dirac Operators with Strongly Singular Potentials," , p. 185, *Spectral Theory and Differential Equations*, *Lecture Notes in Mathematics*, No. 448, ed. A. Dold, B. Echmann, and W.N. Everitt (Springer-Verlag, N.Y. (1975)

[11] J. Von Neumann, *Mathematical Foundation of Quantum Mechanics*, (Princeton Univ. Press, 1955)

[12] Near the origin[9] for $\Gamma < \frac{1}{4}$ ($\nu > 0$), $\chi \approx \frac{1}{2} \pi \rho^{\frac{1}{2}} \csc \pi \nu \left[ -\frac{(\frac{1}{2}\rho)^\nu}{\Gamma(1+\nu)} - \frac{(\frac{1}{2}\rho)^{\nu+2}}{\Gamma(2+\nu)} + \frac{(\frac{1}{2}\rho)^{-\nu}}{\Gamma(1-\nu)} + \frac{(\frac{1}{2}\rho)^{-\nu+2}}{\Gamma(2-\nu)} \right]$;

and for $\Gamma = \frac{1}{4}$, $\chi = \rho^{1/2} [\psi(1) - \ln(\frac{1}{2}\rho) + ...]$

[13] A. Sommerfeld, *Atombau and Spektrallinien*, *Vol. II*, (F. Vieweg u. Sohn, Braunschweig, 1939), considers behavior of Schrödinger and Dirac equations at $r = \infty$.

[14] P.A.M. Dirac, *The Principles of Quantum Mechanics*, (Oxford Univ. Press, London, 1958), considers the Coulomb potential in Schrödinger's equation. Extending his procedure, note that if $\psi \to r^{-1+s}$ near the origin, then $\lim_{r_0 \to 0} [\int^{r_0} d^3r \, |\psi|^2] = 0$ if $s > -\frac{1}{2}$ s or $\infty$ if $\leq -\frac{1}{2}$; however $\int^{r_0} d^3r \nabla^2 \psi = -4\pi(1-s) \lim_{r_0 \to 0} r_0^s$. For Coulomb potential, $s = 0$ and $\nabla^2 \psi = -4\pi \delta(\vec{r})$; if $-1 < s < 0$, $\nabla^2 \psi = 4\pi(1-s)(1+s)^{-1} r^{1+s} \partial_r \delta(\vec{r})$; etc. Thus if the wave equation is not to contain an attractive $\delta$-function or its derivatives as a source (inhomogeneous term), it is required that $s > 0$ even though the function is square integrable for $s > -\frac{1}{2}$.